\documentclass[12pt,preprint]{aastex}

\newcommand{\kms}{\mbox{km s$^{-1}$}}
\newcommand{\hmol}{H$_{2}$~}
\newcommand{\nh}{N$_{\rm{H I}}$}
\newcommand{\rv}{$v_{rad}$}
\newcommand{\opdep}{$\tau(\lambda)$}
\newcommand{\eratio}{$E_{abs}/E_{em}$~}

\usepackage{url}
\usepackage{multirow}
\makeatletter

\newcommand{\Rmnum}[1]{\expandafter\@slowromancap\romannumeral #1@}
\makeatother

\begin{document}

\title{\emph{HST}/COS SPECTRA OF DF TAU AND V4046 SGR: FIRST DETECTION OF MOLECULAR HYDROGEN
       ABSORPTION AGAINST THE Ly$\alpha$ EMISSION LINE \altaffilmark{1}}

\altaffiltext{1}{Based on observations made with the NASA/ESA \emph{Hubble Space Telescope}, obtained from the data
    archive at the Space Telescope Science Institute. STScI is operated by the Association of Universities for
    Research in Astronomy, Inc. under NASA contract NAS 5-26555.}


\author{Hao Yang}
\affil{JILA, University of Colorado and NIST, Boulder, CO 80309-0440} 
\email{haoyang@jilau1.colorado.edu}

\author{Jeffrey L. Linsky}
\affil{JILA, University of Colorado and NIST, Boulder, CO 80309-0440} 
\email{jlinsky@jilau1.colorado.edu}

\author{Kevin France}
\affil{CASA, University of Colorado, Boulder, CO 80309-0389} 
\email{kevin.france@colorado.edu}

\begin{abstract}

We report the first detection of molecular hydrogen (H$_{2}$) absorption in the Lyman-$\alpha$ 
emission line profiles of two classical T Tauri stars (CTTSs), DF Tau and 
V4046 Sgr, observed by \emph{HST}/COS. This absorption is the energy source for many of the Lyman-band H$_{2}$ 
fluorescent lines commonly seen in the far-ultraviolet spectra of CTTSs. 
We find that the absorbed energy in the H$_{2}$ pumping transitions from a portion of the Lyman-$\alpha$ line 
significantly differ from the amount of energy in the resulting fluorescent emission. 
By assuming additional absorption in the \ion{H}{1} Lyman-$\alpha$ 
profile along our light of sight, we can correct the H$_{2}$ absorption/emission ratios so that they are close 
to unity. The required \ion{H}{1} absorption for DF Tau is at a velocity close to the radial velocity of the star, 
consistent with \ion{H}{1} absorption in the edge-on disk and interstellar medium. For V4046 Sgr, a nearly face-on system, 
the required absorption is between $+100$ \kms\ and $+290$ \kms, most likely resulting from \ion{H}{1} gas in the 
accretion columns falling onto the star.

\end{abstract}


\section{INTRODUCTION}

Molecular hydrogen (H$_{2}$) emission lines are commonly observed in the far-ultraviolet (FUV) spectra of 
classical T Tauri stars (CTTSs) \citep{Herczeg2002, Herczeg2004, Herczeg2006, Ardila2002}. 
\citet{brown1981} first detected these lines from T Tau using \emph{IUE}. 
As suggested by detections of identical lines in the sunspot spectrum \citep{jordan1977}, 
these fluorescent H$_{2}$ lines are thought to be photoexcited from the ground electronic state to the B (or C) 
electronic state primarily by coincidence with hydrogen Lyman-$\alpha$ but also by other strong atomic emission lines 
(e.g., \ion{C}{2}, \ion{C}{4}, and \ion{O}{6} ) in the UV. 

The H$_{2}$ fluorescence may arise from various locations in protostellar systems. In their analysis of 
\emph{HST} STIS E140M spectra of 6 pre-main sequence (PMS) stars, \citet{Herczeg2006} found that blueshifted H$_{2}$ lines of 
RU Lupi, T Tau, and DG Tau are likely formed in outflows, while the H$_{2}$ lines of 
TW Hya, DF Tau, and V836 Tau show no radial velocity shifts from the photospheric lines and are likely formed in warm
($\sim$2500 K) surfaces of their circumstellar disks. 
Stars such as T Tau also show spatially extended \hmol fluorescent emission in associated nebulosity, 
which is likely pumped by local shocks and outflows rather than stellar Lyman-$\alpha$ emission \citep{Walter2003}.
For the diskless counterparts of CTTSs, the naked T Tauri stars (NTTSs), H$_{2}$ features 
are not seen in their FUV spectra \citep[][ Yang et al. 2011, in preparation]{Ingleby2009},
indicating that \hmol fluorescent emission requires the presence of \hmol gas close to the central star. 
Studying the \hmol fluorescent emission therefore provides valuable information on the physical properties of protoplanetary 
disks, which are 90\% composed of \hmol gas.

While the \hmol fluorescent lines have been studied in a number of CTTSs, the pumping transitions had not been observed in 
absorption against the Lyman-$\alpha$ emission line. 
In this Letter, we present new FUV spectroscopy of two CTTSs, DF Tau and V4046 Sgr, for which we detect such absorption 
for the first time as a result of the very low noise and high throughput of the Cosmic Origins Spectrograph (COS) on \emph{HST}.

\section{OBSERVATIONS AND DATA REDUCTION} 

We observed DF Tau (RA = 04:27:02.795, DEC = 57:12:35.38) and V4046 Sgr (RA = 18:14:10.466, DEC = -32:47:34.50) with the 
COS \citep{Dixon2010,Osterman2011}
on January 11 and April 27, 2010, respectively. COS is a high-throughput, moderate-resolution UV 
spectrograph installed on the \emph{HST} in May 2009. 
During our \emph{HST} GTO program 11533, we used both the G130M and G160M gratings of the COS FUV channel 
to cover the 1136 \AA--1796 \AA\ region.
Since there is a small gap ($\sim$15 \AA\ in wavelength coverage) between the two segments of the COS detector, we observed 
each star at 4 central wavelength settings for both gratings to provide continuous spectral coverage and minimize any fixed-pattern noise.
The total exposure time for each star was about 10,000 sec during 4 \emph{HST} orbits. The spectral resolution was approximately $17,000$--$18000$, 
with extended wings in the line-spread function. The extended wings are induced by polishing errors on the \emph{HST} primary and
secondary mirrors \citep[see][]{ghavamian2010}.

We reduced the DF Tau and V4046 Sgr spectra using the COS calibration pipeline, CALCOS\footnotemark (v2.12, March 19, 2010), and combined them
with a custom IDL coaddition routine described by \citet{danforth2010}.
In Figure~\ref{fig01}, we show two portions of the coadded FUV spectra for both stars as examples. 
To match the atomic and molecular emission lines with laboratory wavelengths, we corrected for the radial velocity of DF Tau (+15 \kms),
but found that an additional correction of +8 \kms\ is needed because of inaccuracies in the CALCOS wavelength solution. 
The radial velocity of V4046 Sgr is close to 0, and the wavelength solution is accurate.

\footnotetext{We refer the reader to the Cycle 18 COS Instrument Handbook for more details: 
      \url{http://www.stsci.edu/hst/cos/documents/handbooks/current/cos_cover.html}.}

\section{DF TAU AND V4046 SGR}

DF Tau is a binary system consisting of two early M stars separated by 0.1\arcsec\ \citep{schaefer2006}.
Its disk is inclined by 80--85$^\circ$ \citep{cmj2001,Ardila2002}, i.e., the disk is seen nearly edge-on.
The distance to DF Tau is generally adopted as $140$ pc, the distance to the Taurus Molecular Cloud.
\citet{Herczeg2006} analyzed in detail
the \hmol fluorescent lines of DF Tau observed with the STIS E140M grating. 
\citet{herczeg2008} estimated a visual extinction ($A_{V}$) of 0.6 mag and accretion 
rate in the range of $2.3$--$4.6 \times 10^{-8} M_{\odot}$ yr$^{-1}$ for DF Tau.

V4046 Sgr is a spectroscopic binary with a separation of 9 $R_{\odot}$ and an orbital period of 2.42 days \citep{stempels2004}.
The pair consists of a K7V and a K5V star \citep{quast2000}.
At a distance of $\sim 72$ pc (Torres et al. 2008), it is an isolated system and the extinction is practically $A_{V} = 0.0$ mag \citep{stempels2004}.
The circumbinary disk is inclined at 35$^{\circ}$ \citep{quast2000,kastner2008}, somewhat face-on. V4046 Sgr may be 
a member of the $\beta$ Pic Moving group and could be as old as 12 Myr (Ortega et al. 2002). \citet{jensen1997} found that it has little excess 
emission in the near-infrared (near-IR) wavelengths but a large excess emission at longer wavelengths, suggesting that optically thick dust 
in the regions of the disk close to the star has been cleared out. We know of no previous detailed study of \hmol emission from V4046 Sgr.

Besides the \hmol line emission studied in this work, the FUV continuum, which is also detected in DF Tau and V4046 Sgr, will be analyzed in detail 
by K. France et al. (2011, submitted).

\section{ANALYSIS \& RESULTS}

The FUV spectra of DF Tau and V4046 Sgr are dominated by \hmol fluorescent lines (see Figure~\ref{fig01}).
Below 1200 \AA, there are also many Werner-band \hmol lines, which will be described in a separate paper (Yang et al. in preparation). 
In Figure~\ref{fig02}, we show the Lyman-$\alpha$ profiles of DF Tau and V4046 Sgr. The apparent absorption
features seen against the Lyman-$\alpha$ emission line have depths much greater than the noise at those 
wavelengths and coincide in wavelength with the Lyman-band \hmol pumping transitions \citep[see Table 3 of][]{Herczeg2006}. 
We first mask out the absorption features and fit 5th or 6th order polynomial curves to the 
Lyman-$\alpha$ profiles. To measure the amounts of absorbed energy, we integrate the area between the fitted curves and the
observed spectra. The uncertainties in such measurements are mainly caused by the somewhat subjective determinations of the Lyman-$\alpha$ 
profiles without absorption.
Some features result from the blended absorption of two or three pumping transitions, and the absorbed energy for these features represents
the total absorption of the transitions.
For DF Tau, 5 features are measured that correspond to 9 pumping transitions: 1218.52 \AA\ + 1218.57 \AA, 1219.09 \AA\ + 1219.10 \AA\ + 1219.15,
1219.37 \AA\ + 1219.48 \AA, 1219.74 \AA, and 1220.18 \AA. 
For V4046 Sgr, 8 features are measured that correspond to 11 pumping transitions: 1212.43 \AA\ + 1212.54 \AA, 1213.36 \AA, 1213.68 \AA, 
1217.03 \AA\ + 1217.04 \AA, 1217.20 \AA, 1217.41 \AA, 1217.64 \AA, and 1217.90 \AA\ + 1217.98 \AA.

We detect between 2 and 19 fluorescent \hmol lines in the progressions produced by each pumping transition observed 
in absorption against the Lyman-$\alpha$ lines.
The \hmol fluorescent lines are identified based on the line list of \citet{abgrall1993}. To measure the line fluxes, 
we used a custom IDL fitting procedure \citep{France2010BD} that convolves 
a Gaussian profile with the COS line spread function (LSF) to fit the observed \hmol line profiles.
The convolution of a Gaussian profile with the COS LSF only changes 
the shape of the profile but not the total line flux. The uncertainties in the line fluxes are generally less than 5 \% for
unblended lines, indicative of the high signal-to-noise of the data, typically $\geq 50$ for the H$_{2}$ lines. 
We next convert the line fluxes in each progression to the total energy emitted from the pumped upper level.
Each \hmol line in a given progression yields an estimate of 
the total energy in the upper level from the line theoretical branching ratios. We average the 
estimated total energy emitted from each upper level using only the strong unblended lines, and their
standard deviations are $\le$ 15\% of the mean values.
The emission from each upper level is corrected for its dissociation probability as calculated by \citet{abgrall2000}. 
The dissociation probability is typically zero or only a few percent.
The absorption and emission fluxes for DF Tau are also corrected for extinction using $A_{V}=$ 0.6 mag and the \citet{cardelli1989} 
extinction law.

For each absorption feature observed against the Lyman-$\alpha$ emission line, 
we have estimated the absorbed energy ($E_{abs}$) and the emitted energy ($E_{em}$)
from the corresponding \hmol upper level. If there are no additional sources of absorption or emission, 
then the global average of \eratio should be unity.
We begin the analysis by assuming that \eratio along our line of sight should be close to unity, but the measurements show otherwise. 
We do find that the three features in the Lyman-$\alpha$ blue wing for V4046 Sgr have \eratio within a factor
of two of unity (1.38, 0.99, 0.56), but the \eratio ratios at longer wavelengths are smaller than one by factors of 4--60 
(see the fourth column of Table~\ref{table01}). 
We therefore propose that additional Lyman-$\alpha$  absorption in the line of sight between the location where \hmol is pumped and the observer has
reduced the observed absorbed energy in the pumping line and thus the \eratio ratios.
To model the additional hydrogen absorption, we calculate Lyman-$\alpha$ absorption profiles using the Voigt function for a range of hydrogen column 
densities (\nh) and radial velocities (\rv). Then for each combination of \nh\ and \rv, we calculate the optical depths, \opdep, at the 
wavelengths of the pumping lines. We correct the observed \eratio ratios by multiplying by $e^{-\tau(\lambda)}$. For V4046 Sgr, the absorption features
at 1217.03 \AA\ and 1217.20 \AA\ are not used. The 1217.03 \AA\ feature is close to the line center of Lyman-$\alpha$, and the measurement 
is greatly affected by the absorption in the line center. The 1217.20 \AA\ progression does not have enough strong unblended emission lines 
to provide an accurate estimate of the emission energy.

In Figure~\ref{fig03}, we show for DF Tau and V4046 Sgr the combinations of \nh\ and \rv\ that can correct the \eratio ratios to be close to unity. 
For DF Tau, if we assume that the extinction is all due to interstellar dust, we can convert $A_{V}$ = 0.6 mag to a total hydrogen column density of
log($N_{H}$) $= 21.03$ according to the relation in \citet{predehl1995}.
Note that this value assumes a standard gas-to-dust ratio for the interstellar medium.
We mark in the top panel of Figure \ref{fig03} the corresponding \nh\ values assuming 100\% and 50\% neutral hydrogen. 
\citet{Herczeg2006} estimated for DF Tau the absorption against the red wing of the Lyman-$\alpha$ emission line and measured a log(\nh) of 20.75. 
This value is close to the hydrogen column density converted from $A_{V}$ with 50\% neutral content and represents the neutral hydrogen in both the
interstellar medium along the line of sight and possibly the edge-on disk of the system. 
As shown in Figure~\ref{fig03}, within reasonable ranges of \nh, the additional absorption required to 
bring all of the \eratio ratios close to unity for DF Tau
requires radial velocities close to zero. Given that the uncertainty in $A_{V}$ could be as large as 0.4 mag, and the ionization fraction 
and the gas-to-dust ratio in the disk are unknown, 
we think that the absorption is likely caused by a combination of interstellar medium and neutral hydrogen in the edge-on disk, though
absorption by the \ion{H}{1} columns in stellar winds or accretion columns can not be completely ruled out.

For V4046 Sgr, which suffers negligible continuum extinction, the absorption required has a radial velocity between 
100 and 290 \kms, as shown in the bottom panel of Figure~\ref{fig03}. 
If $A_{V}$ is close to zero \citep{stempels2004}, then the radial velocity is close to 290 \kms. For this case,  we list in Table~\ref{table01} 
the calculated optical depth for the additional absorption as well as the measured and corrected \eratio for the absorption features detected in V4046 Sgr.
This is consistent with a scenario in which \ion{H}{1} in the accretion columns is absorbing the red wings of the Lyman-$\alpha$ emission line 
in this system, which is oriented somewhat face-on.

\section{DISCUSSION}

\begin{figure}[ht]
  \begin{center}
    \includegraphics[scale=0.75]{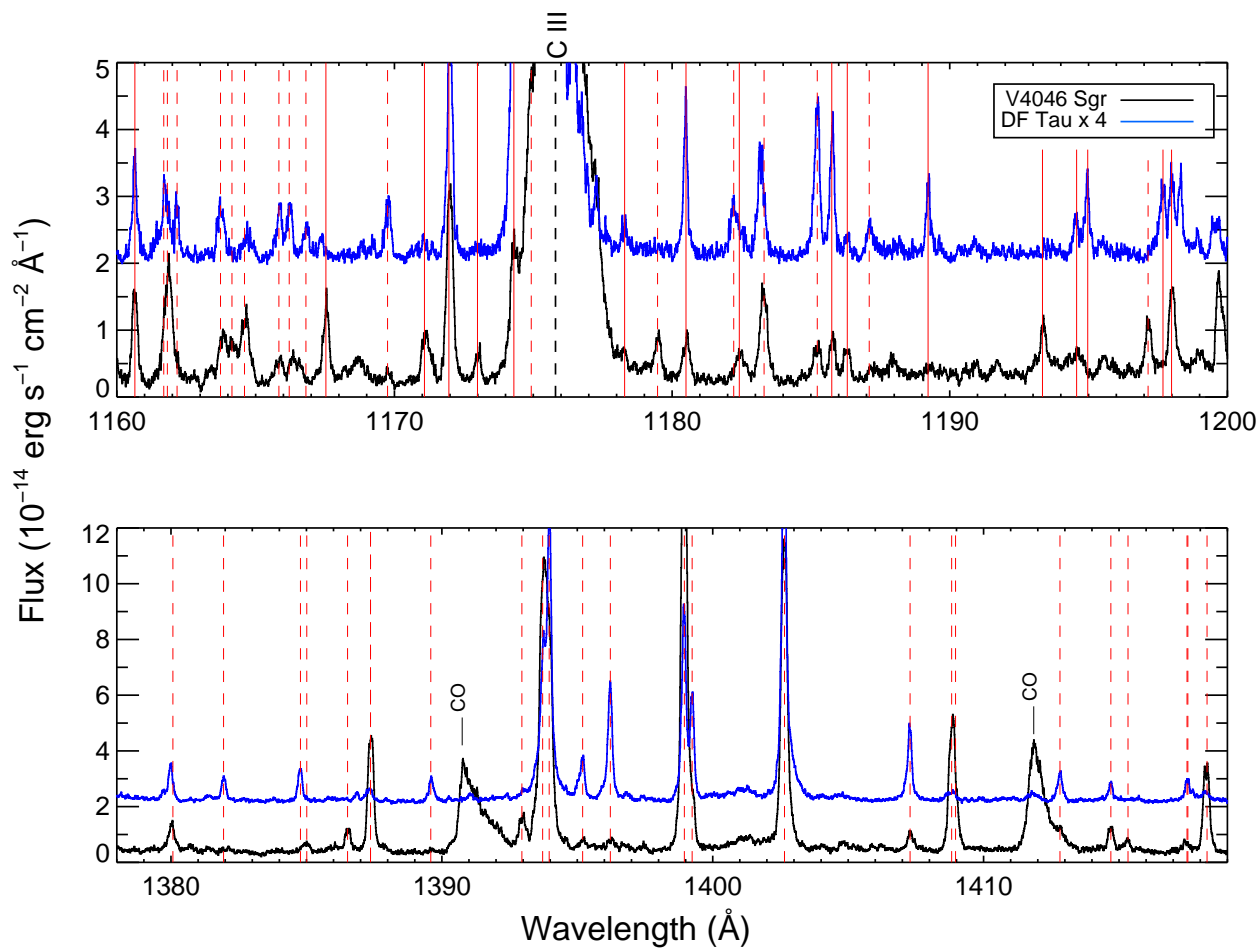}
       \caption{Two portions of the COS G130M spectra of DF Tau and V4046 Sgr. For display purposes,
      the DF Tau spectrum is scaled up by a factor of 4 and offset by $+2 \times 10^{-14}$ erg s$^{-1}$ cm$^{-2}$ \AA$^{-1}$.
      The red dashed vertical lines identify the fluorescent Lyman-band H$_{2}$ emission lines and the red solid vertical lines 
      identify the fluorescent Werner-band H$_{2}$ lines. The strong emission feature around 1176 \AA\ is the \ion{C}{3} multiplet. Two carbon 
      monoxide (CO) emission features are also marked in the lower panel and will be presented by K. France et al. (2011, in preparation). }
           \label{fig01}
              \end{center}
    \end{figure}

\begin{figure}[ht]
  \begin{center}
    \includegraphics[scale=0.75]{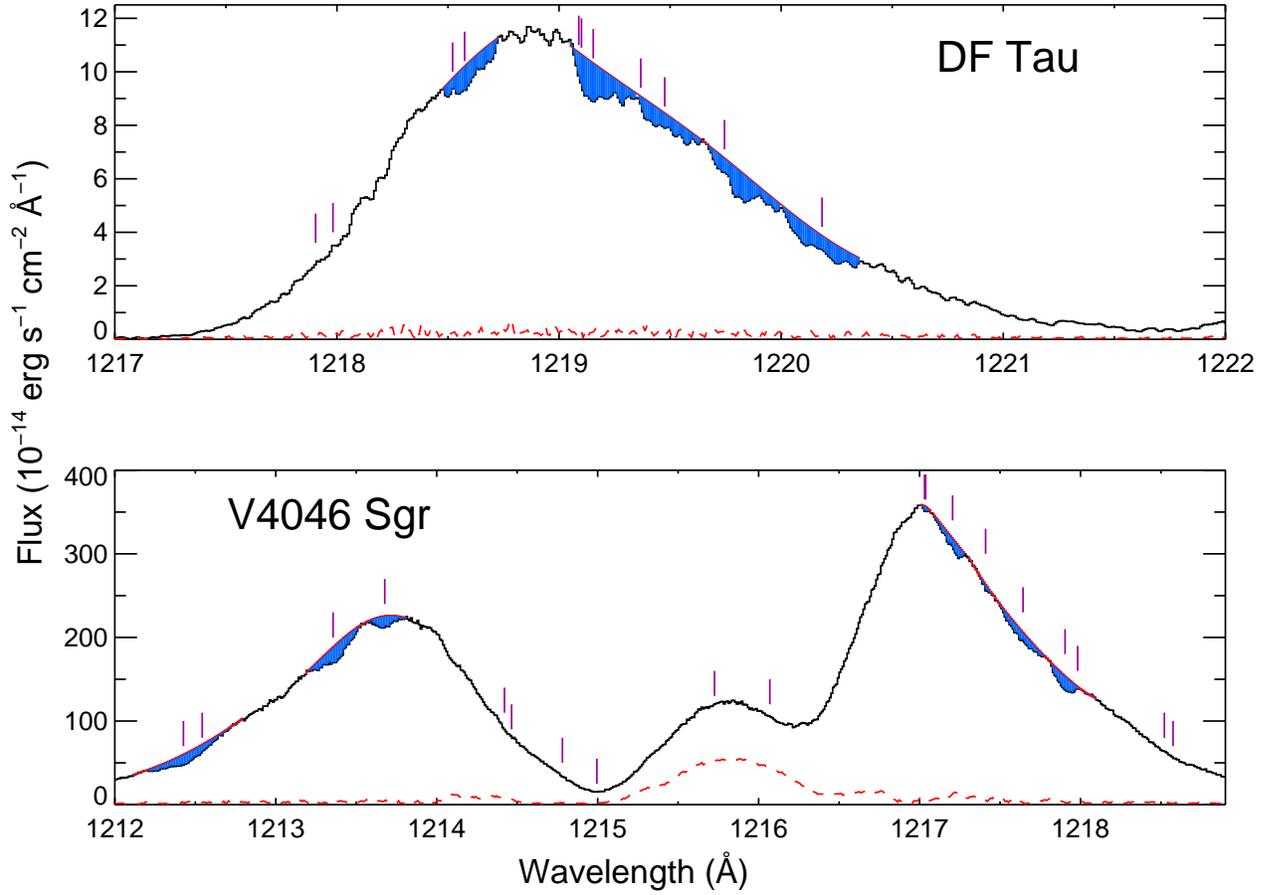}
       \caption{The Lyman-$\alpha$ emission profiles of DF Tau (\emph{top panel}) and V4046 Sgr (\emph{bottom panel}). 
       The red dashed lines indicate the uncertainty levels, and the vertical ticks mark
       the wavelengths of coincident Lyman-band H$_{2}$ transitions. The blue areas indicate our estimates of 
       the absorption in the \hmol pumping transitions. In the \emph{bottom panel}, the emission bump centered around 
       1215.7 \AA\ is geocoronal emission and is not from the star.}
           \label{fig02}
              \end{center}
    \end{figure}

\begin{figure}[ht]
  \begin{center}
    \includegraphics[scale=0.75]{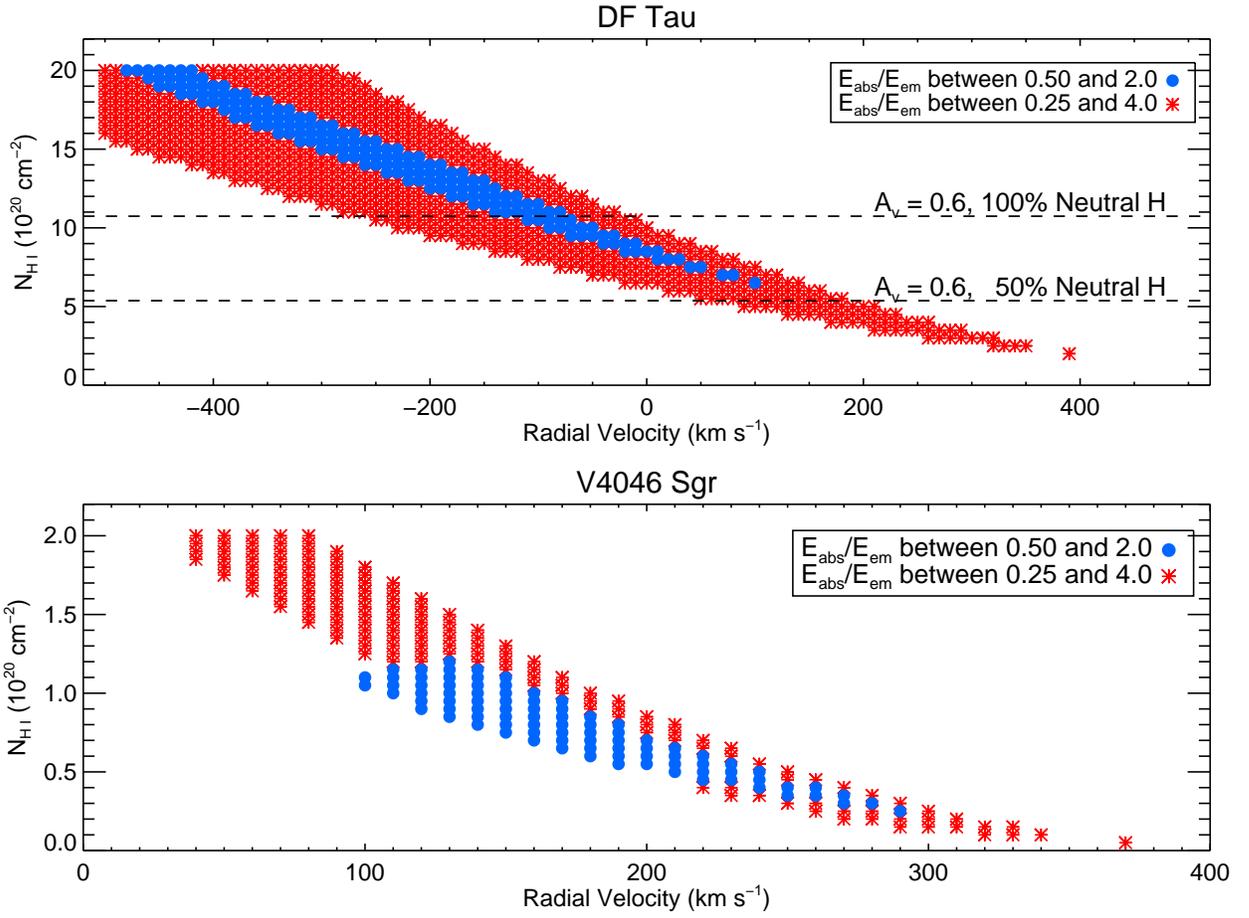}
       \caption{H$_{2}$ absorption/emission ratios of DF Tau (\emph{top panel}) and V4046 Sgr (\emph{bottom panel}),
       corrected for additional Lyman-$\alpha$ absorption for a range of \ion{H}{1} column 
        densities and radial velocities. }
           \label{fig03}
              \end{center}
    \end{figure}

Thanks to the excellent sensitivity and low background of COS, we were able to detect for the first time \hmol absorption against 
the Lyman-$\alpha$ emission line profiles in two CTTSs. Because of the large aperture of COS (2.5\arcsec), the center of 
the Lyman-$\alpha$ line is filled with geocoronal Lyman-$\alpha$ emission and not usable (see Figure~\ref{fig01}). The STIS E140M spectrum of DF Tau
reported by \citet{Herczeg2006} shows that the line center and blue wing of Lyman-$\alpha$ are completely absorbed. Interstellar
absorption must be responsible for the disappearance of the line center, because we see many \hmol lines pumped
by transitions coincident with the center of Lyman-$\alpha$. On the other hand, we detect only a few \hmol lines pumped 
by the transitions blueward of line center, suggesting that the blue-wing emission of Lyman-$\alpha$ has been absorbed by the stellar
wind before the blue-wing radiation reaches the molecular gas in the disk. Since the disk of DF Tau is viewed nearly 
edge-on, the stellar wind must be present near the stellar equator to absorb the blue wing of the Lyman-$\alpha$ emission line.

\begin{figure}[ht]
  \begin{center}
    \includegraphics[scale=0.45]{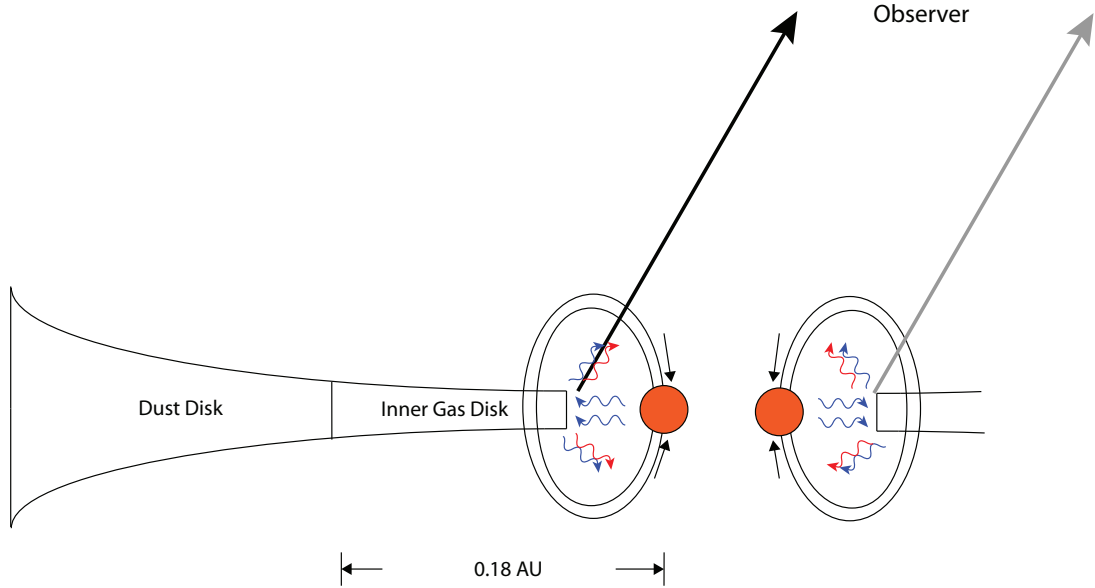}
       \caption{A schematic cartoon of the V4046 Sgr system, which is not drawn to scale. 
       The blue wavy arrows represent stellar and reflected Lyman-$\alpha$ photons, and the red wavy arrows 
       represent the fluorescent \hmol photons. The dust disk begins at 0.18 AU from the star, according
       to \citet{jensen1997}. The short black arrows indicate the flow direction along the accretion funnels.
       The long black and gray arrows point to our line of sight, which is at $\sim$35$^\circ$ 
       with respect to the rotation axis of the disk. The long black arrow indicates that there is more reflected light from 
       the inner edge of the gas disk that faces toward our line of sight than that from the opposite side, 
       which is indicated by the long gray arrow, because of absorption through the disk along this line of sight.}
           \label{fig04}
              \end{center}
    \end{figure}

Understanding the geometry of the V4046 Sgr system requires more detailed consideration. 
The stellar wind must be weak for this somewhat face-on system since the blue wing of Lyman-$\alpha$ is not totally absorbed. 
Our results show that there is additional absorption in the red wing of Lyman-$\alpha$ that could be explained by accretion 
with velocities that are at least 100 \kms\ and are likely as large as 290 \kms. We envision a model in which a portion of the Lyman-$\alpha$
emission, likely formed near the accretion shocks, is reflected to the observer by neutral hydrogen in the inner disk. 
A schematic cartoon of this model is shown in Figure~\ref{fig04}.
The \hmol pumping and fluorescence occurs where these Lyman-$\alpha$ photons are present in the inner disk, as described by the ``thick disk''
model in \citet{Herczeg2004}. The reflected Lyman-$\alpha$
emission line, including the \hmol absorption at the pumping wavelengths, is then absorbed by infalling neutral hydrogen in large 
accretion funnels in our line of sight. 


Accretion of gas from circumstellar disks onto CTTSs is generally thought to be controlled by 
stellar magnetic fields \citep{bouvier2007}. Strong magnetic fields of a few kilogauss
\citep[][Yang et al. 2010, submitted]{cmj2007} truncate the circumstellar gas disk at a few stellar radii 
and direct the accretion funnels onto the star near the magnetic poles.
From models of the spectral energy distribution (SED) of V4046 Sgr, \citet{jensen1997} found that dust in the inner regions is cleared out 
to about 0.18 AU, which is 38.6 $R_{\odot}$. At this distance, the stellar magnetic fields are not strong enough to interact 
efficiently with the disk, and, more importantly, the temperature of molecular gas at the surface of the dusty disk may not be high enough
for the hydrogen molecules to be electronically excited by the Lyman-$\alpha$ photons. For this transitional disk system, we think that 
both the fluorescent emission and accretion columns likely originate from the inner molecular gas disk \citep{Muzerolle2003L,kastner2008}
that is closer to the central star than the dust disk. The absence of any significant differences between the radial velocities of the
star, pumping lines and fluorescent lines is consistent with the fluorescent \hmol gas lying in an inner gas disk seen nearly face-on. 
In this picture as shown in Figure~\ref{fig04}, we expect to see more reflected light from the inner edge of the ``thick disk'' 
(cf. Figure $8$ in \citealt{Herczeg2004}) facing toward our line of sight than from the opposite side where the disk absorbs light along
the line of sight to the observer. Therefore, the reflected Lyman-$\alpha$
emission line summed over all viewing angles toward both stars will show absorption mostly by downflowing neutral hydrogen in the
accretion columns (seen along the long black line in Figure~\ref{fig04}) rather than upflowing neutral hydrogen 
(seen along the long gray line in Figure~\ref{fig04}).

\citet{guenther2007pro} modeled the accretion shocks on V4046 Sgr, and their best-fit model of the X-ray 
observations yields an accretion rate of $3\times10^{-11} M_{\odot}$ yr$^{-1}$ and a maximum infall velocity of 535 \kms\
where the accretion gas strikes the stellar surface. Since the \ion{H}{1} absorption velocity is between $+100$ and $+290$ \kms, the absorbing 
\ion{H}{1} gas is located between 2.3 $R_*$ and 4.4 $R_*$ above the accretion shock (cf. Eq. $1$ in \citealt{calvet1998}) 
if we assume that the accretion columns have a constant cross-sectional area and the infalling gas sees only gravitational forces.


\acknowledgments 

This work was support by NASA grants NNX08AC146 and NAS5-98043 to the University of Colorado at Boulder.
The authors would like to thank the COS Science Team for help with data reduction and useful discussions.

\bibliographystyle{apj}

\begin{deluxetable}{cccccc}
  \tablewidth{0pt}
  \tablecaption{ Lyman-$\alpha$ Absorption Features of V4046 Sgr. \label{table01}}
\tabletypesize{\scriptsize}
\tablehead{
\colhead{Wavelengths (\AA)} & \colhead{Pumping Transitions} & \colhead{$P_{Dis}$\tablenotemark{a}} 
&\colhead{Measured \eratio} &\colhead{Optical Depth\tablenotemark{b}} &\colhead{New \eratio}
}
\startdata
1212.43 and 1214.54 & 1-1 P(11) and 1-1 R(12) & 0.0 and 0.0     & 1.377    & 0.055  & 1.456  \\
1213.36             & 3-1 P(14)               & 0.015           & 0.999    & 0.087  & 1.090  \\
1213.68             & 4-2 R(12)               & 0.050           & 0.563    & 0.106  & 0.626  \\
1217.41             & 4-0 P(19)               & 0.417           & 0.047    & 3.505  & 1.568  \\
1217.64             & 0-2 R(1)                & 0.0             & 0.098    & 1.700  & 0.537  \\
1217.90 and 1219.98 & 2-1 P(13) and 3-0 P(18) & 0.002 and 0.189 & 0.226    & 0.894  & 0.553  \\

\enddata
\tablenotetext{a}{Dissociation probability from upper level, calculated from \citet{abgrall1993}.}
\tablenotetext{b}{The optical depth is calculated for the additional absorption with $\rm{N_{H I}}$ = $2.5 \times 10^{19}$ cm$^{-2}$ and
         $v_{rad} = 290$ \kms.}
\end{deluxetable}

\clearpage

\clearpage

\clearpage

\clearpage

\end{document}